\documentclass[a4paper,
11pt,
unpublishedssss
]{quantumarticle}

\pdfoutput=1
\usepackage[utf8]{inputenc}
\usepackage[english]{babel}
\usepackage[T1]{fontenc}
\usepackage{amsmath}
\usepackage{amsfonts}
\usepackage{hyperref}
\usepackage{tikz}
\usepackage{lipsum}
\usepackage{longtable}
\usepackage{algorithm}
\usepackage{algpseudocode}
\usepackage{comment}
\usepackage{booktabs}

\algrenewcommand\algorithmicrequire{\textbf{Input:}}
\algrenewcommand\algorithmicensure{\textbf{Output:}}

\begin{document}
\title{A Rubik's Cube inspired approach to Clifford synthesis}
\author{Ning Bao}
\affiliation{Northeastern University, Boston, MA USA}
\affiliation{Brookhaven National Laboratory, Upton, NY USA}
\orcid{0000-0002-3296-1039}
\email{n.bao@northeastern.edu}

\author{Gavin S. Hartnett}
\email{hartnett@rand.org}
\orcid{0000-0002-6814-1809}
\affiliation{RAND Corporation, Santa Monica, CA USA}

\maketitle

\begin{abstract}
The problem of decomposing an arbitrary Clifford element into a sequence of Clifford gates is known as Clifford synthesis. 
Drawing inspiration from similarities between this and the famous Rubik's Cube problem, we develop a machine learning approach for Clifford synthesis based on learning an approximation to the distance to the identity. 
This approach is probabilistic and computationally intensive. However, when a decomposition is successfully found, it often involves fewer gates than existing synthesis algorithms. 
Additionally, our approach is much more flexible than existing algorithms in that arbitrary gate sets, device topologies, and gate fidelities may incorporated, thus allowing for the approach to be tailored to a specific device.
\end{abstract}

\section{Introduction}
The field of quantum computing has seen significant progress over the past three decades. Algorithms for factoring \cite{shor1999polynomial}, quantum simulation \cite{georgescu2014quantum}, and solving linear systems \cite{harrow2009quantum} hold great promise for the ability of quantum computers to revolutionize mathematics and physics.

Many of these algorithms, however, require very large numbers of qubits and quantum gates to outperform classical computers with modern memories and processing speed. 
The largest among the current generation of quantum computers have a few hundred qubits, with typical two-qubit entangling gate error rates around 1\% \cite{google2023suppressing, kim2023scalable}. 
Although these devices are too small and noisy to execute the most powerful quantum algorithms, when augmented with various error mitigation and resilience methods, they have nevertheless been sufficient to demonstrate \emph{quantum supremacy} \cite{arute2019quantum, morvan2023phase}, the ability of quantum computers on carefully chosen problem to outperform any existent (and hopefully any, full-stop) classical competitor, thus providing experimental proof of principle of the short-term returns for quantum algorithms. 
While the problems considered for these quantum supremacy demonstrations were largely selected for their computational complexity properties and not for their importance to science or mathematics more broadly, the point remains that genuine quantum advantage has been demonstrated, and will eventually extend to problems of more practical interest.

In this current noisy intermediate-scale quantum (NISQ) era of quantum computing \cite{preskill2018quantum}, quantum circuit compilation, or the streamlining of quantum circuits down to their shortest and most resource-efficient form, is of paramount importance. 
Quite simply, a circuit with gate-depth 150 might be too noisy to be worth running on NISQ devices, but one with gate-depth 80 might not be. 
Consequently, relatively microscopic optimizations that improve the actual performance of circuits executed on real devices, but which are irrelevant in the complexity-theoretic, big-O sense, are quite important for finding novel quantum algorithms that can demonstrate near-term quantum advantage.

Human intuition for quantum circuit optimization, however, is quite poor at scaling as system sizes become larger and more complex. 
It is therefore a natural question to ask whether machine learning techniques can do better in this regard. 
While human designers may be constrained by classical programming intuition or by lamppost effects overemphasizing existing algorithmic approaches, it is possible that machines will not share one or both of these particular limitations in the context of quantum circuit design. 
Indeed, work in this area already exists, in the work of \cite{fosel2021quantum} for reinforcement learning (RL) approaches to quantum control, \cite{moro2021quantum} approaches to optimization of single qubit gates, and \cite{weiden2023improving} for unitary synthesis using seed synthesis techniques.

In this work, we will focus on the specific problem of Clifford circuit optimization, as opposed to that of a generic quantum circuit. 
There has been much work done on this in the past, see for example \cite{aaronson2004improved, bravyi2021hadamard, bravyi2021clifford, bravyi20226}, though this work has not settled on optimal 
Clifford circuit implementations of a given Clifford unitary for arbitrary number of qubits $n$ (However, optimal synthesis algorithms have been obtained for $n=2,3,4,5,6$ in for example \cite{bravyi2021clifford}.). 
We will seek to find more efficient Clifford circuits for given Clifford unitaries than those that the existing methods are able to generate.

In some sense, the Clifford problem is highly analogous to solving a generalized Rubik's Cube: there are well-defined sets of operations (or moves), a clear notion of success, and a finite group structure. 
Despite this, obvious loss functions such as the number of matched faces on the Rubik's Cube or the trace distance for Clifford unitaries do not provide useful learning signals for machine learning-based approaches. 
For example, in the course of solving the Rubik's problem it is typical to make apparently destructive moves that seem to undo previous progress in order to reach the solution. 
In the context of the Rubik's Cube problem, recent work has demonstrated how to use deep learning to develop useful heuristics, or guidance functions, which can be used to identify a sequence of steps that will solve the cube from a general scrambled state \cite{johnson2019stepwise, johnson2018solving, johnson2021solving, agostinelli2019solving}.
We will adapt some of these techniques for the Clifford synthesis problem. 
Our is not the first work on unitary synthesis to draw inspiration from the analogy with the Rubik's Cube. Ref.~\cite{bravyi2021clifford} developed an exhaustive computational approach capable of finding optimal synthesis for Cliffords which scales to $n=6$ qubits, and \cite{zhang2020topological} adapted the DeepCubeA approach for solving the Rubik's Cube \cite{agostinelli2019solving} for the problem of synthesizing topological Fibonacci anyons.

Our approach will be twofold; first, we will introduce a graph structure in which edges representing certain single Clifford gates connect pairs of vertices representing Clifford unitaries that are related by those gates (this graph is known as the Cayley graph).
We will simply argue that a Djikstra search \cite{dijkstra1959note} on this graph would be sufficient to find the optimal Clifford implementation for each unitary, with appropriate adjustment of the edge weights. 
The obvious limitation to this approach, e.g. the size of the graph at hand, will also be discussed, with some comments on how it could be ameliorated.

The second approach will use a learned guidance function approach to find relatively short Clifford circuit implementations of the given unitary. 
We find that when this approach succeeds at finding a decomposition, it often beats or matches existing, non-optimized heuristic approaches in terms of the gate count (and thus, circuit depth). 
However, this approach is probabalistic in nature and is not guaranteed to succeed, and thus is best thought of as a way to occasionally improve upon the decomposition furnished by existing approaches. 
We have made a Python implementation of this second approach publicly available here: \url{https://github.com/gshartnett/rubiks-clifford-synthesis}.
 
Lastly, it should be noted that deterministic solutions exist for synthesis or approximate synthesis of any quantum circuit, a la Solovay-Kitaev \cite{kitaev1997quantum}, Dawson-Nielson \cite{dawson2005solovay}, and Kliuchnikov \cite{kliuchnikov2013synthesis}. 
There has also been recent work on unitary gate synthesis of non-Clifford gates, as in \cite{patel2021robust}. 
Slightly further afield, there is work relating complexity to geometry, as in \cite{Nielsen_2006}.

\section{Rubik's Cube}

We begin with a brief review of the Rubik's Cube and its group-theoretic structure, as used in \cite{johnson2019stepwise, johnson2018solving, johnson2021solving}. 
The standard Rubik's Cube is a $3 \times 3 \times 3$ cube with 6 faces and $6 \times 9 = 54$ ``facets''. 
Each facet is assigned one of six colors, with nine facets of each color. 
By rotating the faces of the cube the positions of these facets may be changed, and the goal is to apply moves to the cube so that each face consists of facets of a single color. 

It is convenient to adopt the convention that the orientation of the cube is held fixed as the faces are rotated - for example that the center facet facing the user is always white. 
Within this convention, the  distinct cube configurations, or states, can be enumerated by assigning each non-center facet a number 1 through 48 and considering permutations of these numbers. 
Importantly, not all permutations correspond to valid cube states reachable from the solved state through allowed moves. 
Each move may be associated with a particular permutation, and the set of valid cube states consists of all possible compositions of these. 
There is thus a one-to-one correspondence between the set of all valid cube states and the subgroup of the symmetric group $S_{48}$ (the group of all permutations of 48 elements) generated by the move set permutations. 
This subgroup is termed the Rubik's group and is denoted $G_{\text{Rubik's}}$. 
The solved cube is associated with the identity permutation $e$. In this group-theoretic description, the problem of solving the Rubik's Cube amounts to identifying a sequence of permutation moves $x_1, x_2, ..., x_K$ which compose to give the inverse of current state $x$, i.e., $x \left( x_1 x_2 ... x_K \right) = e$. 
Each of the permutation moves must lie in the move set, the allowed moves of the cube. 
The two commonly used move sets are the half-turn metric, which allows rotations of a face by $90^{\circ}$, $180^{\circ}$, or $270^{\circ}$, and the quarter-turn metric, which allows rotations by $90^{\circ}$ or $270^{\circ}$ only.

The Rubik's group is huge, containing 43,252,003,274,489,856,000 distinct elements. Despite this fact, it has been proven that any two cube states may be connected through a short sequence of moves. 
The ``God's number'' $\ell_{\text{God}}$ is defined as the maximum of the minimal number of moves needed to connect any two states. 
The precise value of the God's number depends on the move set: it is 20 using the half-turn metric and 26 using the quarter-turn metric \cite{rokicki2014diameter, rokicki2023gods}.

From an algorithmic standpoint, it is useful to describe the Rubik's Cube problem as a graph traversal problem, where the graph in question is the Cayley graph.
The nodes in this graph correspond to the cube states (group elements), and two nodes are connected by a directed edge if and only if there is a move in the move set connecting them.
The graph will be undirected if the inverse of every move in the move set is also in the move set (as is the case for both the half-turn and quarter-turn metrics). 
Moreover, the graph is regular, with the degree of each node given by the size of the move set. 
The problem is then: given an arbitrary starting node, find a path (preferably a geodesic) from that node to the solved state node. 
The God's number corresponds to the maximum length geodesic between the solved state node and any other node. 
It turns out that this is equivalent to the diameter of the Rubik's graph (the largest geodesic distance between any two nodes). 

\section{Clifford synthesis}
The Clifford group is the normalizer of the Pauli group. 
In this work we use the notation Cl($n$) to denote the finite-dimensional $n$-qubit Clifford group with the overall phase, corresponding to the $U(1)$ center, projected out. 
Clifford elements may be represented as tableaus, ${2n \times (2n + 1)}$ binary-valued matrices of the form
\cite{aaronson2004improved}
\begin{equation}
    T = 
    \begin{pmatrix}
    \tilde{X} & \tilde{Z} & \tilde{p} \\
    X & Z  & p
    \end{pmatrix} 
    \,,
\end{equation}
where $X, \tilde{X}, Z, \tilde{Z}$ are themselves $n \times n$ binary matrices constrained so that the square part of $T$, denoted $T'$ (i.e., with the phase bit column omitted) satisfies the symplectic condition, $(T')^T \Omega_n T' = \Omega_n$, where 
$$
\Omega_n = 
\begin{pmatrix} 0 & I_n \\ I_n & 0 \end{pmatrix}
\,,
$$ 
$I_n$ is the $n\times n$ identity matrix, and where $\tilde{p}, p$ are each length-$n$ binary vectors. 
The rows of $T$ indicate both the stabilizer group generators (rows $n+1$ to $2n$) and the destabilizer generators (rows $1$ to $n$). 
The overall phase $(\pm 1)$ of the (de)stabilizer for a given row in the tableau is encoded via the phase bit vectors.
With this definition, the Clifford group can be identified with the Cartesian product of the space of $2n \times 2n$ binary symplectic matrices with the space of length $2n$ binary vectors, $\text{Cl}(n) = \text{Sp}(2n, \mathbb{F}_2) \times \mathbb{F}_2^{2n}$, and thus the size can be seen to grow rapidly (more precisely, doubly exponentially) with $n$ \cite{bravyi20226} 
\footnote{In Sec.~\ref{sec:results} we present numerical results for a slightly simpler version of the problem where the phase bits are dropped, in which case the factor of $2^{2n}$ should be omitted.} 
\begin{equation}
    \label{eq:Cldim}
    \text{dim(Cl(}n\text{))} = 2^{n^2 + 2n} \prod_{i=1}^n \left( 2^{2i} - 1 \right) \,.
\end{equation}

The problem of Clifford synthesis is: given an element of the $n$-qubit Clifford group Cl($n$), find a quantum circuit with gates drawn from some generating set that will implement that element. 
This is also referred to as Clifford decomposition. The most widely used generating set is the set of $H$ and $S$ gates for each qubit, as well as the set of all CNOT gates for every pair of qubits, but of course other generating sets are possible. 
There are also two versions of the problem which differ in how Clifford elements should be represented: either as $2^n \times 2^n$ unitary matrices, or as $2n \times (2n + 1)$ binary-valued tableaus. 
The former makes clear that this is a special case of the more general problem of unitary synthesis, while the latter is convenient because the tableau structure directly enforces the condition that the unitary be a member of the Clifford group.

The analogy between Clifford synthesis and the Rubik's Cube problem can now be established. 
The group structure is already evident - problem states are Clifford group elements which are naturally represented as tableaus. 
Group composition corresponds to matrix multiplication (mod 2) of the $2n \times 2n$ square component of the tableaus (with the appropriate accounting used to keep track of the phase bits), or as matrix multiplication if the group elements are represented as $2^n \times 2^n$ unitary matrices. 
The solved state corresponds to the identity tableau/matrix. For each choice of move set the Clifford group can be endowed with a graph, with two nodes connected by an edge if the corresponding tableaus are connected via a move from the move set. 
The move set corresponds to a set of generating gates, such as the set of all single-qubit $H$, $S$ gates as well as two-qubit CNOT gates. 
(Note that because $S$ is not Hermitian, this choice of move set will result in a directed graph.). 
From a group theoretic standpoint, the move set corresponds to a set of generators for the Clifford group, and the associated graph is the Cayley graph.
To facilitate comparison with existing Clifford synthesis approaches implemented in Qiskit \cite{cross2018ibm} we will use the following gate set: all single qubit $X, Y, Z, H, S, S^{\dagger}$ gates, a $CNOT$ gate for each ordered pair of qubits, and a $SWAP$ gate for each unordered pair of qubits. 
Table~\ref{table:analogy} compares how the scale of the Rubik's and Clifford problems compare.

Framing Clifford synthesis in this way is useful because it allows the immediate application of graph traversal algorithms, including general graph algorithms such as Dijkstra or Bellman-Ford, as well as more specialized algorithms designed for Rubik's Cube problems, which we explore below. 
It also allows for notions like the God's number to be established. In this case the Clifford God's number is the maximum length geodesic in the graph connecting the identity tableau to any other tableau. 
Using the move set described below and weighting all gates equally, the God's number for the $n=2$ Clifford group can be found to be 8 via exhaustive computer search.
\footnote{While it is clear that the God's number and diameter are the same for regular, unweighted graphs, changing either of these will differentiate the two concepts, and so we will use the God's number as the figure of merit going forward.}

\begin{table*}
\centering
\begin{tabular}{ | l | l | l | l | l | l | l |  }
 \hline
 & Rubik's & Cl(2) & Cl(3) & Cl(4) & Cl(5) & Cl(6) \\
 \hline
 Group dim. & $4.3 \times 10^{19}$ & 11,520 & 92,897,280 & $1.21 \times 10^{13}$ & $2.54 \times 10^{19}$ & $8.52 \times 10^{26}$ \\ \hline
Num. moves & 12/18 & 15 & 27 & 42 & 60 & 81 \\ \hline
Feature dim. & 54 & 20 & 42 & 72 & 110 & 156 \\ \hline
\end{tabular}
\caption{\label{table:analogy} Analogy between the Rubik's group and the $n$-qubit Clifford group Cl($n$). 
The dimension of the Clifford group is obtained using Eq.~\ref{eq:Cldim}. The number of moves depends on how the problem is being defined. 
For the Rubik's Cube, the two common definitions are the half-turn metric, in which rotations by $90^{\circ}$, $180^{\circ}$, and $270^{\circ}$ are allowed, and the quarter-turn metric, in which only rotations by $90^{\circ}$ or $270^{\circ}$ (or equivalently, left and right quarter turns) are allowed. 
The move set for the Clifford group corresponds to the single qubit gates $X, Y, Z, H, S, S^{\dagger}$, the directional two-qubit $CNOT$ gate, and the symmetric two-qubit $SWAP$ gate. 
An all-to-all qubit connectivity is assumed. 
The features of the Rubik's Cube correspond to all 54 facets, and the features of the tableau are the matrix entries.}
\end{table*}

Not all gates should be counted equally in circuit synthesis. 
Standard circuit identities, such as the decomposition of a SWAP gate into 3 alternating CNOTs, should be incorporated. 
Additionally, hardware-specific fidelities should also be incorporated. 
For example, it is generally the case that a CNOT, implemented on real NISQ hardware, will have a lower fidelity than single qubit gates, but the precise quantification of this comparison will vary across device, and even within a device due to inhomogeneities, as well as with time due to system fluctuations. 
Our graph-based approach easily accomodates these issues through an appropriate assignment of edge weights. 
As each edge corresponds to an allowed move, represented by a Clifford gate, a physically-motivated edge weight assignment is based on the relative gate fidelities. 
The overall (and irrelevant) scale of the weights may be fixed by setting $w = 1$ for edges corresponding to some reference gate $g_*$, for example a Hadamard gate. 
The weights of all other gate-edges can then be taken to be determined by $f(g)^w = f(g_*)$, where $f(g)$ is the fidelity of gate $g$. 
In other words, a weight $w$ gate is equivalent, in terms of fidelity, to $w$ copies of the reference gate $g_*$. 
This weight assignment differs from previous approaches for Clifford synthesis which aim to minimize only the number of two-qubit entangling gates, with no consideration given to single-qubit gates. 

\section{Learned guidance function}
Denote the geodesic distance on the Cayley graph as $d(\cdot, \cdot)$. 
The distance between any tableau $T$ and the identity tableau $T_{\text{id}}$, $d_{\text{id}}(T) := d(T, T_{\text{id}})$ is of special significance: this is the minimal distance  (or equivalently, the weighted number of gates) required to synthesize the tableau $T$. 
It is straightforward to show that the problem of Clifford synthesis can be solved in $O(M \ell_{\text{God}})$ steps given an oracle capable of returning $d_{\text{id}}(T)$, where $M$ denotes the size of the move set and $\ell_{\text{God}}$ the God's number. 
Given an arbitrary tableau, the distance to the identity for each possible move can be retrieved with $M$ calls to the oracle. 
Next, the move that leads to the greatest reduction in distance to the identity will be applied (randomly breaking ties if they arise). 
By repeating this greedy strategy after each move, the identity tableau is guaranteed to be found after no more than $\ell_{\text{God}}$ moves, resulting in a worst-case time complexity of $O(M \ell_{\text{God}})$. 

However, in practice, such an oracle will not be available, and the above algorithm will require that the distances be computed, for example using Djikstra's shortest path algorithm. 
The worst-case time complexity of this is ${O(|E| + |V|\log|V|)}$ for a graph with vertex set $V$ and edge set $E$. The move set graph is $M$-regular, and so $|E| = M |V|$, resulting in $O(|V|\log|V|)$. 
Unfortunately, this scaling is impractical for even moderately large circuit widths $n$ given the fact that ${|V| = \text{dim(Cl(}n\text{))}} = O(2^{n^2})$. 
\footnote{Note that this analysis is naive and could likely be improved. In particular, the regularity of the graph has not been used. Practical implementations can also make use of the fact that algorithms like Djikstra do not require the full graph to be stored in memory.}

Although the poor scaling renders the above algorithm impractical, it does serve to motivate the heuristic approach we develop here. 
The central idea is to avoid applying Djikstra's algorithm to a doubly exponentially-large graph and to instead develop an approximation to the distance to the identity, $g(T) \approx d_{\text{id}}(T)$. 
In particular, we will model $g$ as a neural network. 
Provided that $g(T)$ is a sufficiently good approximation, calls to Djikstra's algorithm may be replaced with feed-forward evaluation of $g$ which will require a number of evaluations which grows only polynomially in $n$. 
Of course, this does not account for the time complexity required to learn a sufficiently good approximation; the exponential improvement will come at the cost of the algorithm becoming probabilistic, as well as the additional complexity of training $g$ to achieve an approximation of sufficient quality. 
Following \cite{johnson2018solving, johnson2019stepwise, johnson2021solving}, we will refer to $g$ as a \textit{learned guidance function} as it will be later used to guide the Clifford synthesis problem. 
The function $g$ could equally well be called a \textit{heuristic}, for example in the context of A* search and related graph traversal algorithms.

\subsection{Model and training details \label{sec:model}}
We chose to model $g: \mathbb{R}^{2n, 2n + 1} \mapsto \mathbb{R}$ as a feed-forward neural network with all-to-all connectivity. 
(Note that valid tableaus are binary symplectic matrices, but the neural network does not require the entries to be binary nor the symplectic condition to be satisfied.) 
In particular, a network with 3 hidden layers with dimensions $[32, 16, 4]$ is used. 
Each layer is followed by a non-linear activation (the logistic sigmoid is used for all but the final layer which applies the exponential function). 
We note that we did not attempt to optimize this model architecture using meta-learning techniques, and that doing so may yield additional improvement, at the cost of the meta-learning overhead.

The weights $\theta$ of the network will be chosen to encourage the guidance function to be a good approximation of the distance to the identity. This will be accomplished by minimizing a suitable loss function over a training dataset: ${\theta_* = \text{argmin}_{\theta} \, \mathcal{L}(\theta)}$. In formulating the loss function, an important observation is the fact that an ideal guidance function need not match the distance pointwise, instead it need only satisfy the weaker condition that it leads to the same ordinal ranking among the move set,
\begin{equation}
    \label{eq:idealguidancefunction}
    d_{\text{id}}(T_1) < d_{\text{id}}(T_2) \, \text{   iff   } g(T_1) < g(T_2) \,,
\end{equation}
for any two tableaus $T_1, T_2$ connected by a move (i.e., the corresponding nodes are connected by an edge in the graph).

In practice we can only hope to learn a guidance function which satisfies Eq.~\ref{eq:idealguidancefunction} a fraction of the time. 
A natural loss function that penalizes violations is the negative Pearson correlation coefficient averaged over the training dataset, 
\begin{equation}
    \mathcal{L}_{\text{Pearson}}(\theta) = -\frac{1}{N_B} \sum_{i=1}^{N_B} r_{D_{\text{id}}, g} \,.
\end{equation}
Here $N_B$ is the number of batches in the training set and $r_{D_{\text{id}}, g} \in [-1,1]$ is the Pearson correlation coefficient for two equal length sequences $\{D_{\text{id}}(T_i)\}_{i=1}^B$, $\{g(T_i)\}_{i=1}^B$:
\begin{equation}
    r_{x,y} := \frac{ \sum_{i=1}^K (x_i - \bar{x}) (y_i - \bar{y}) }{ \sqrt{\sum_{i=1}^K (x_i - \bar{x})^2 \sum_{i=1}^K (y_i - \bar{y})^2  }} \,.
\end{equation}
This choice of loss function differs from prior approaches using learned guidance functions to solve Rubik's Cubes. 
Those have instead attempted to solve the more restrictive problem of directly modeling the distance (i.e., number of moves away from the solved cube state) rather than the weaker task of modeling the distance up to a monotonic transformation. 
Of course, the main motivation for adopting the current approach is to avoid the need to calculate the actual distance $d_{\text{id}}$; therefore in the above $D_{\text{id}}$ represents an efficiently calculable upper bound for the true distance which we will discuss momentarily.

The loss function is minimized using a gradient-based optimization. 
The training data is generated by randomly sampling sequences of length-$L$ Clifford gates to generate tuples of the form $(T, \tilde{D}_{\text{id}})$, where $T$ is the Clifford tableau formed by the sampled gates and $\tilde{D}_{\text{id}}$ is both the weighted number of gates and an upper bound to the distance to the identity. 
Each gate is sampled uniformly from the move set, and the sequence length is sampled uniformly from the range 1 to $L_{\text{max}}$, with $L_{\text{max}}$ a hyper-parameter. 
This process may be thought of as a random walk on the problem graph. 
The length of the generated sequence controls the extent to which the walk explores the entire Clifford group. 
Ideally this length would be at least as large as $\ell_{\text{God}}$ to ensure that the walk is capable of reaching all tableaus, but unfortunately we are not aware of any bounds of $\ell_{\text{God}}$ for the Clifford synthesis problem. 
Therefore, we considered different scalings of the length of the walk (size of the generated sequence) with $n$. 
Importantly, note that this random walk will not in general correspond to uniformly sampling the Clifford group, as for example in \cite{bravyi2021hadamard}. 
Lastly, the batch size $B$ and number of batches (epochs) $N_B$ are left as training hyper-parameters. 

\subsection{Greedy algorithm}
A learned guidance function enables a straightforward greedy algorithm, Algorithm~\ref{alg:greedy}, towards Clifford synthesis (dubbed hillclimbing in \cite{johnson2019stepwise, johnson2018solving, johnson2021solving}). 
Given an arbitrary starting Clifford, at each step the learned guidance function is evaluated for every move in the move set and the move with the smallest value of the guidance function is made. 
Ties are broken randomly. 

\begin{algorithm}[H]
\caption{Greedy algorithm}\label{alg:greedy}
\begin{algorithmic}
\Require learned guidance function $g$
\Require initial tableau $x \in \text{Cl}(n)$
\Ensure gate decomposition $x = x_0 \, x_1 \, ... \, x_{K-1}$
\State $i \gets 0$
\State $y \gets x^{-1}$
\While{$y \neq \text{IdentityTableau}(n)$}
    \State $x_i = \text{argmin}_{x' \in \text{MoveSet}} g(y \, x') $ %
    \State $y \gets y \, x_i$
    \State $i \gets i + 1$
\EndWhile
\end{algorithmic}
\end{algorithm}

\subsection{Beam search algorithm}
We also considered beam search as a second algorithm for using the learned guidance function to synthesize Cliffords. 
Beam search can be considered both as a generalization of the greedy algorithm as well as a restriction of breadth-first-search where at each step only the top-$w$ ranked nodes (ranked according to their guidance function values) are retained. 
If $w = 1$, then beam search reduces to the greedy algorithm, and if $w$ is infinite then breadth-first search is recovered. 
Algorithm~\ref{alg:beamsearch} contains pseudocode for a simple implementation of beam search. 
Here $\mathcal{N}$(node) denotes the neighborhood of a node, isSolution(node) returns True if the node is a solution, False if not, and TopRankedBeams returns the top-$w$ ranked nodes in the beam according to their guidance function value. 
The algorithm as written returns the solution node; the full decomposition $x = x_0 x_1 ... x_{K-1}$ can be recovered using for example linked lists, so that each node can be linked to its predecessor.
\begin{algorithm}[H]
\caption{Beam search algorithm}\label{alg:beamsearch}
\begin{algorithmic}
\Require beam width parameter $w$
\Require learned guidance function $g$
\Require initial tableau $x \in \text{Cl}(n)$
\Ensure gate decomposition $x = x_0 \, x_1 \, ... \, x_{K-1}$
\State $\text{Beam} = \{ x \}$
\State $\text{Visited} = \{ x \}$
\While{$|\text{Beam}| > 0 $}
	\For{node in Beam}
		\For{neighbor in $\mathcal{N}$(node)}
		    \If{isSolution(neighbor)}
		    	\State \Return neighbor
		    \EndIf
		    \If{neighbor $\notin$ Visited}
			    \State Beam = Beam $\cup$ \{neighbor\}
		    \EndIf
		\EndFor
		\State Visited = Visited $\cup$ $\mathcal{N}$(node)
    \EndFor
    \State Beam = TopRankedBeams(Beam, $g$, $w$)
\EndWhile
\end{algorithmic}
\end{algorithm}

\section{Results \label{sec:results}}
We considered the Clifford synthesis problem for a range of circuit widths, $n=3, 4, ..., 12$. 
In each case a separate learned guidance function was trained and used to guide the two graph traversal algorithms considered, greedy search, (Algorithm~\ref{alg:greedy}), and beam search with a width $w=3$, (Algorithm~\ref{alg:beamsearch}). 
The training details of the guidance function are as follows. 

For each $n$, the learned guidance function was trained by minimizing the Pearson correlation loss over a dataset of randomly sampled Clifford tableaus as described in Sec.~\ref{sec:model}. 
The optimization was carried out using a batch size of $B=2000$ and a fixed number of batches (epochs) $N_B = 1000$. 
The Adam optimizer was used \cite{kingma2014adam}, with learning rate $10^{-3}$. 
We  considered two different scalings of the maximum length of the random walks used to generate the training data, linear: $L_{\text{max}} = 10 \, n$ and log-linear: $L_{\text{max}} = 10 \, n \log_2(n)$ (rounded to the nearest integer). 
Also, for computational convenience we dropped the phase bits, so that each tableau is represented by a binary-valued symplectic matrix. 
Without the phase-bits, tableau composition corresponds to simple matrix multiplication which is easily parallelized.
\footnote{Note, however, that the Aaronson-Gottesman theorem allows Clifford circuits to be simulated more efficiently than matrix multiplication. In particular, matrix multiplication scales as $O(n^{2.37})$ for a $n\times n$ matrix where as the simulation algorithm introduced in \cite{aaronson2004improved} only requires $O(n^2)$ time. Clearly, there is much room for improvement in our implementation, which was mainly guided by the built-in capabilities of modern deep learning libraries (in particular, PyTorch).} 
In order to facilitate a comparison between our method and the built-in Qiskit synthesis functionality, in our experiments we used a gate weighting which corresponds to simply counting the number of CNOT gates: single-qubit gates were given weight 0, CNOT gates were given weight 1, and SWAP gates were given weight 3.

\begin{figure}
    \centering
    \includegraphics[width=0.5\textwidth]{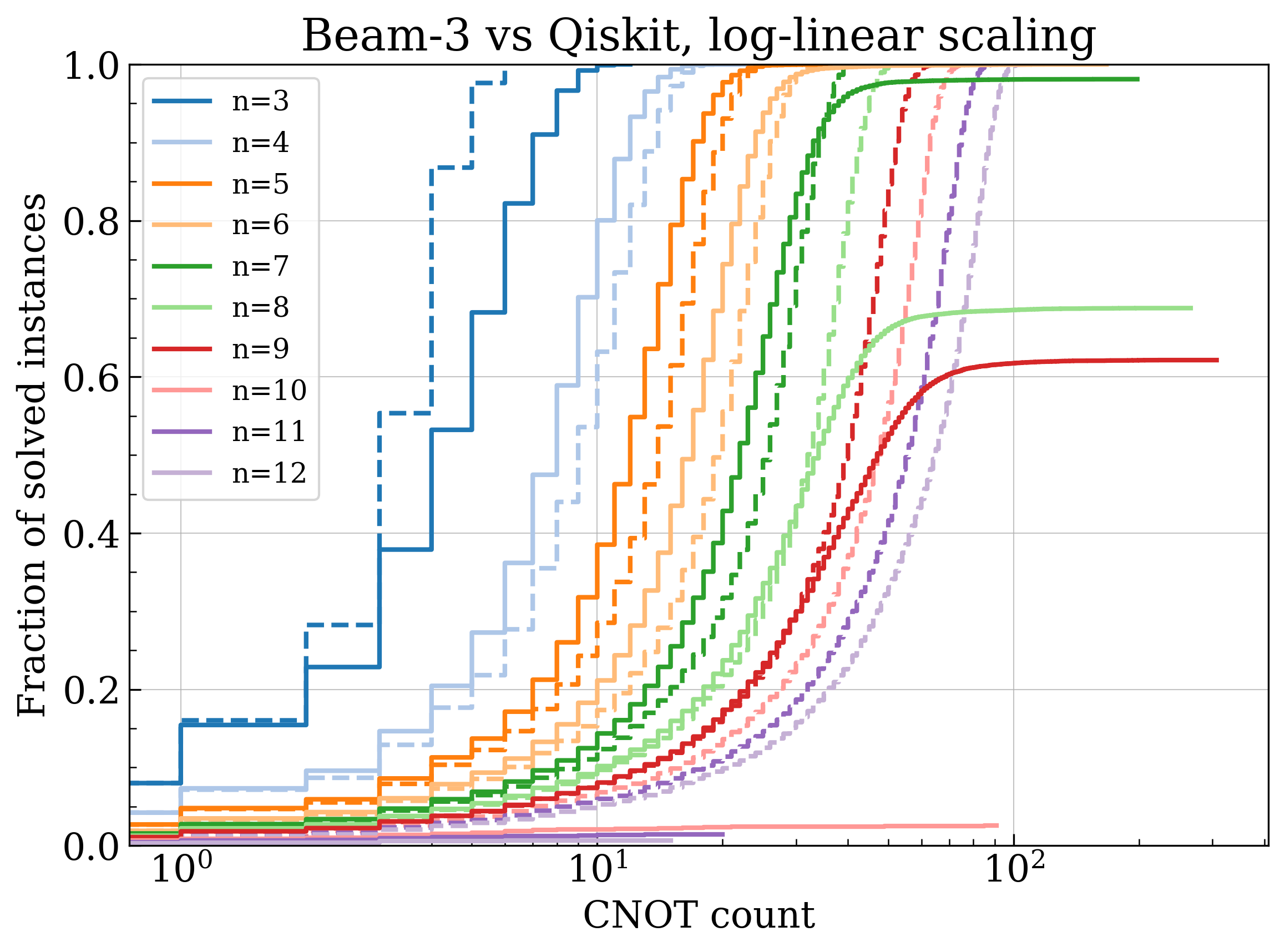}
    \caption{
    The fraction of solved instances as a function of CNOT count for both the beam search algorithm with beam width $w=3$ (solid curve) and the built-in Qiskit method (dashed line).
    }
    \label{fig:beam3_vs_qiskit}
\end{figure}

\begin{table*}[htbp]
  \centering
  \caption{\label{table:comparison_beam_loglinear} Beam search comparison with built-in Qiskit method (log-linear scaling).
  }
  \begin{tabular}{ccccc}
    \toprule
    $n$ & success (\%) & $\ell_{\text{Beam}} < \ell_{\text{Qiskit}}$ (\%) & $\ell_{\text{Beam}} \le \ell_{\text{Qiskit}}$ (\%) & $\frac{\ell_{\text{Qiskit}} - \ell_{\text{Beam}}}{\ell_{\text{Qiskit}}}  $ (\%) \\
    \midrule
    		3 & 100.0 & 0 & 49.0 & N/A \\
                 4 & 100.0 & 52.3 & 89.4 & 25.1 \\    
                 5 & 100.0 & 60.3 & 87.5 & 21.6 \\    
                 6 & 100.0 & 64.4 & 86.1 & 19.6 \\    
                 7 & 98.1 & 65.2 & 83.3 & 17.9 \\    
                 8 & 68.8 & 62.2 & 81.1 & 16.9 \\    
                 9 & 62.2 & 58.5 & 76.6 & 15.6 \\    
                 10 & 2.6 & 17.2 & 96.9 & 15.9 \\    
                 11 & 1.4 & 6.9 & 100.0 & 15.2 \\    
                 12 & 0.7 & 0.0 & 100.0 & N/A \\    
    \bottomrule
  \end{tabular}
\end{table*}

The quality of the quality of the learned guidance function is best judged by how well it guides  graph traversal.
There are two key considerations: first, the fraction of problem instances that can be successfully decomposed, and second, the weighted gate count (path length) of the found decomposition.
To evaluate these, both the greedy and beam search algorithms were applied to Clifford tableaus generated using the same random walk procedure used to train the guidance functions, ensuring that the training and testing distribution over Clifford tableaus were the same. 
Due to the rapid growth of the Clifford group, we can expect that the performance will suffer in the case where these two distributions differ.
As a baseline for comparison, the built-in Qiskit Clifford synthesis function \texttt{synth\_clifford\_full} was also applied to these same problem instances.
At the time of writing, the Qiskit function incorporates three previously published algorithms \cite{aaronson2004improved, bravyi2021hadamard, bravyi2021clifford}). 
To evaluate the greedy algorithm, 20,000 Clifford tableaus were generated, and the algorithm was terminated if it failed to reach the identity tableau after 1000 steps. 
\footnote{
Exceptions to this are: in the case of the linearly scaled beam search algorithm for $n \ge 8$, and the log-linearly scaled beam search algorithm for $n \ge 10$, 5,000 tableaus were generated and the algorithm was terminated after 200 steps.
In these cases, fewer problem instances were considered and the algorithm was terminated after a fewer number of steps due to the long run-times.
}

Among the two algorithms considered, the beam search approach performed best. 
Fig.~\ref{fig:beam3_vs_qiskit} depicts a direct comparison between the guided graph traversal approach using the beam search algorithm and the built-in Qiskit approach for Cliffords sampled using the log-linearly scaled random walk. 
The guided graph traversal algorithm is able to solve nearly all problem instances for $n$ up to 7. 
For $n = 4,5,6$ the cumulative fraction of instances solved by the beam search, as a function of CNOT gates, is greater than or equal to the fraction solved by Qiskit, indicating that in many cases the beam search decomposition utilizes fewer gates. 
For $n=7, 8$ the cumulative fraction solved by beam search is greater than the Qiskit fraction until a threshold CNOT count is reached, beyond which beam search algorithm is unable to find a decomposition.
The performance drops off rapidly as $n$ increases further, indicating that the quality of the guidance function has not kept up with the growth of the problem complexity.
\footnote{
The sizeable drop in performance in going from $n=9$ to $n=10$ is partially attributable to the fact that the beam search algorithm was terminated prematurely, and that with further iterations the guided search would have succeeded in finding a decomposition.
}

Table~\ref{table:comparison_beam_loglinear} contains additional comparative statistics, including the overall fraction of problem instances that can be successfully decomposed (success), the fraction of instances for which the guided graph traversal results in a decomposition with fewer CNOT gates than the Qiskit method, conditioned on a successful decomposition ($\ell_{\text{method}} < \ell_{\text{Qiskit}}$), the fraction of instances for which the guided graph traversal results in a decomposition with fewer \textit{or equal} CNOT gates than the Qiskit method, again conditioning on a successful decomposition ($\ell_{\text{method}} \le \ell_{\text{Qiskit}}$), and lastly the fractional decrease in the number of CNOT gates for problem instances that can be successfully decomposed \textit{and} result in a decomposition using fewer gates than the Qiskit method (${ (\ell_{\text{Qiskit}} - \ell_{\text{method}})/\ell_{\text{Qiskit}}}$). 
(The N/A entries correspond to cases where none of the problem instances met these conditions, and it should also be noted that the Qiskit method is provably optimal with respect to CNOT count for $n=3$.) When the beam search algorithm succeeds in finding a decomposition with fewer CNOT gates than Qiskit, the average reduction in CNOT count (equivalent to the weighted path length of the graph traversal) ranges from 25\% for $n=4$ to 17\% for $n=8$. For $n > 8$ the algorithm increasingly struggles to find decompositions. 

\begin{figure*}
    \centering
    \includegraphics[width=0.95\textwidth]{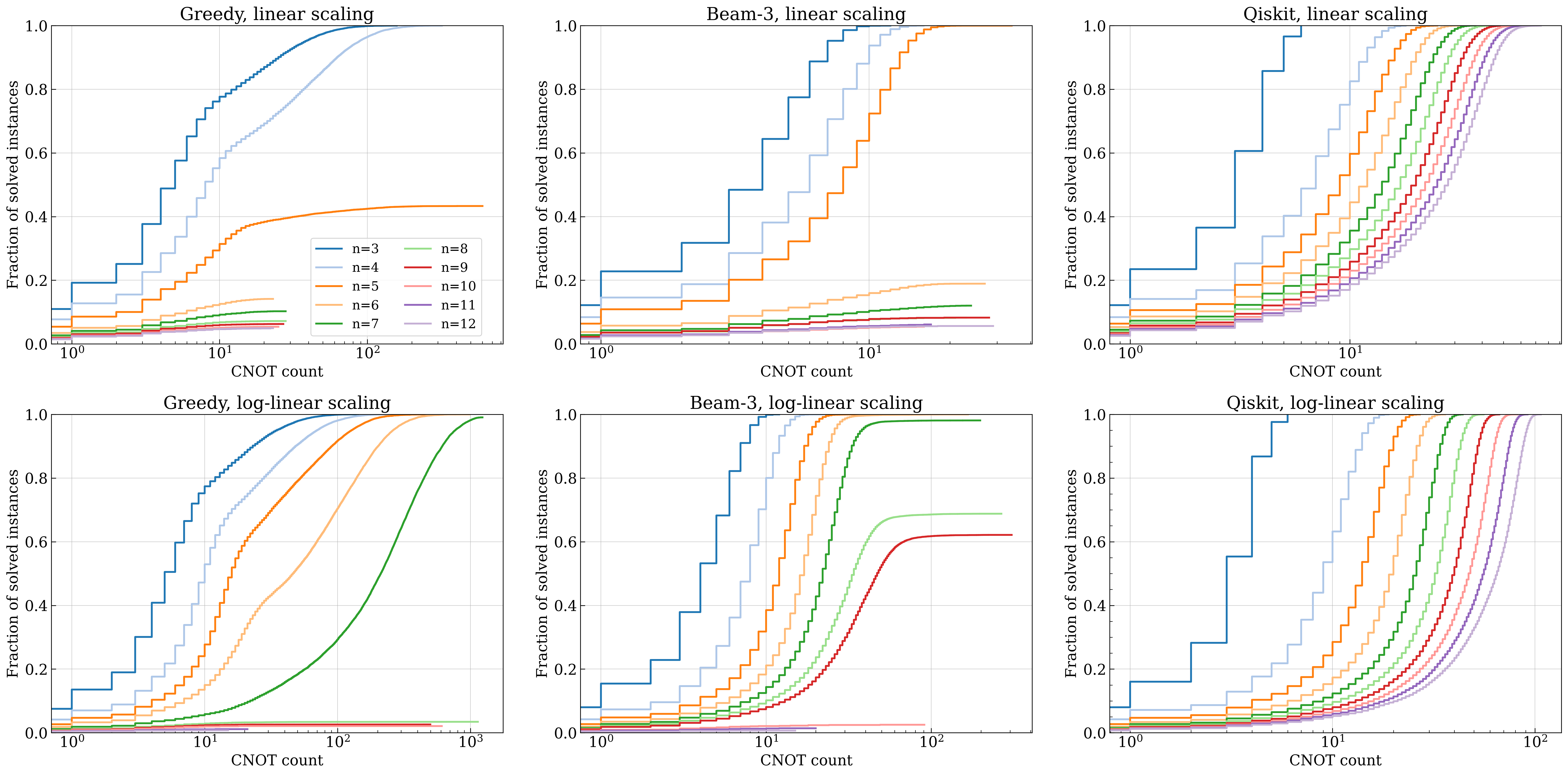}
    \caption{Fraction of problem instances solved. (\textit{Left}): The cumulative fraction of problem instances solved by the greedy algorithm as a function of weighted distance, for a range of circuit widths. (\textit{Center}): The analogous plot for the beam search algorithm with beam width $w=3$. (\textit{Right}): The Qiskit method.}
    \label{fig:cumulative_plot}
\end{figure*}

Fig.~\ref{fig:cumulative_plot} provides a summary of the results for all the considered algorithms; Appendix~\ref{app:supplement} contains further details.
As expected, in all cases the Qiskit approach succeeds in finding a decomposition. 
In contrast, the fraction of instances solved by the guided graph traversal approaches is 100\% for the first few values of $n$, and then decreases as $n$ grows.
The guidance functions trained on log-linearly scaled random walks significantly outperform those trained on the linearly scaled random walks, and the beam search algorithm outperforms the greedy algorithm. 

\section{Discussion}
The problem of Clifford synthesis has many structural similarities with the famous Rubik's Cube problem. 
Inspired by recent machine learning approaches to solving the Rubik's Cube problem, we have developed a learned guidance function approach for the Clifford synthesis problem. 
The drawbacks to our approach is that it only succeeds in finding a decomposition on some fraction of problem instances, requires a computationally intensive training procedure, and in some cases is outperformed by existing algorithms readily available in Qiskit. 
However, when our method is able to find a decomposition, it is often more economical in terms of CNOT gate count than the decompositions produced by existing algorithms. 
For moderate numbers of qubits, $n=4, 5, 6, 7$, the guided beam search method outperforms Qiskit for almost all problems (an optimal decomposition method exists for $n=3$).
Our approach can therefore be used in conjunction with existing approaches to occasionally find improved decompositions of Clifford elements. 
This approach is also much more flexible than existing approaches in that it supports any universal gate set over the Clifford group, as well as arbitrary relative weights (or costs) for each gate. 
This allows the algorithm to be tailored to a given device.

Given that the approach developed here was inspired by approaches used to solve the Rubik's group, it is interesting to compare the scale of the two problems in rough terms. 
By coincidence, the size of the Clifford group for $n=5$ is comparable to the size of the Rubik's group; both contain roughly $10^{19}$ elements. 
The dimensions for $n=6$ and $n=7$ are of the order $10^{27}$ and $10^{35}$, respectively.
Therefore, this work demonstrates that the learned guidance function approach can be applied to much larger problem spaces than previously considered.

There are many ways that our approach could be improved. 
The overall run-time of the guided graph traversal could likely be sped up significantly through better software implementations.
Additionally, we have made no attempt to tune any of the hyper-parameters appearing in the approach. 
One particularly important hyper-parameter is $L_{\text{max}}$, the length of the random walk. 
In particular, we should have $L_{\text{max}} \ge \ell_{\text{God}}$ to ensure that the support of the training distribution is the entire Clifford group  However, the God's number is unknown for general $n$, and if the analogous quantity in the Rubik's Cube problem is to be any guide, it is likely that $\ell_{\text{God}}$ can only be computed via exhaustive computer scan for relatively small values of $n$ \cite{rokicki2014diameter, rokicki2023gods}. 
Absent such exhaustive scans, our approach would benefit from lower bounds on $\ell_{\text{God}}$ if they could be established. From a more pragmatic perspective, the decompositions found by our approach could be improved by applying a simplification pass that applies simple gate identities to reduce the overall gate count.

It would be desirable to more directly encode known properties of the Clifford group into the guidance function. 
For example, the current approach does not enforce that the input matrices are binary-valued or that they are symplectic. 
Nor is any use made of any cosets of the Clifford group, which could potentially be used to significantly reduce the search space.
It would also be interesting to incorporate recent work on fully classifying the Clifford group \cite{grier2022classification}.

Another potential quantum extension to our approach is to upgrade the random walk strategy used to a quantum random walk strategy. 
In a quantum random walk, the adjacency matrix of the graph is taken to be the time independent Hamiltonian guiding the evolution of the quantum state, which is initially fully localized on the initial vertex of the graph. 
For many families of graphs, this would prevent the possibility of the classical random walk from missing the optimal traversal, as the quantum random walk would sample the potential paths in superposition, a la \cite{childs2002example,childs2003exponential}. 
This approach would potentially require a more complete knowledge of the form of the graph, though it is possible that considering only a subgraph would yield usable results.

Because much of the success of our approach in this case is potentially predicated on the discrete structure of the Clifford group, it is unclear whether this approach will work well in the continuous Lie group setting for the group $U(N)$, e.g. for full unitary synthesis. 
In that context, perhaps a more holistic and flexible approach would be appropriate, such as e.g. using a large language model to learn the language and syntax of the languages used for quantum programming. 
Lastly, we note that the problem of Clifford synthesis can also be framed as a simple Markov Decision Process (MDP). 
This framing might be useful for inspiring other algorithmic approaches as well as for making contact with related problems. 

\section{Acknowledgements}
This work grew out of an earlier project in collaboration with Zachary Fisher. 
We also thank Edward Parker and Alvin Moon for their helpful comments on an earlier draft of this manuscript. 
We are grateful to Yuanhang Zhang for helpful discussions, and in particular for bringing our attention to previous work using deep learning and graph traversal algorithms to solve the Rubik's Cube and his related work for unitary synthesis.
N.B. is supported by the DOE Office of Science-ASCR, in particular under the grant Novel Quantum Algorithms from Fast Classical Transforms.

\bibliography{refs}
\bibliographystyle{JHEP}

\onecolumn
\appendix
\newpage
\section{\label{app:supplement} Additional Results}

\begin{figure*}[ht]
    \centering
    \includegraphics[width=0.95\textwidth]{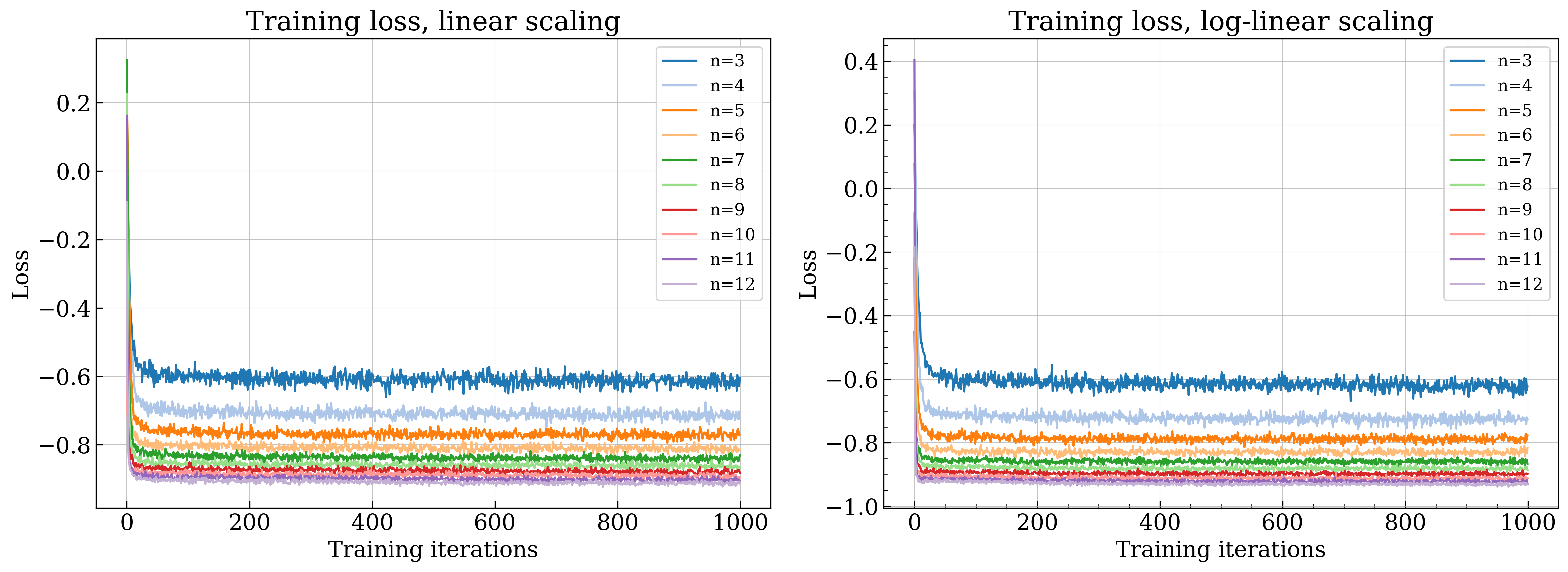}
    \caption{The Pearson correlation loss evaluated at each step in the training process.
    }
    \label{fig:training_loss}
\end{figure*}

\begin{figure}[ht]
    \centering
    \includegraphics[width=0.5\textwidth]{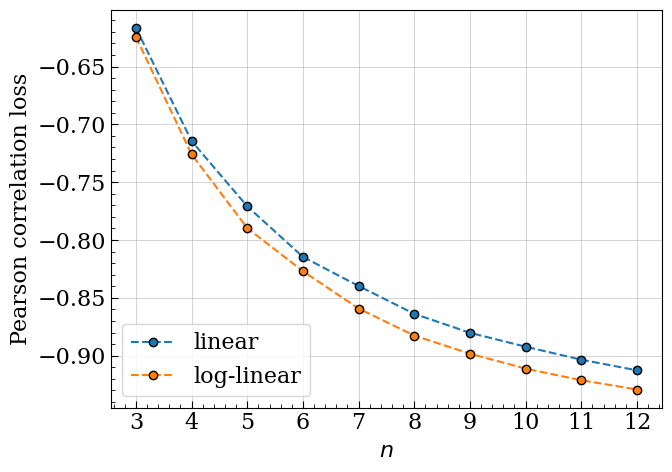}
    \caption{The Pearson correlation loss attained at the end of the training process for the two different scalings of $L_{\text{max}}$. 
    }
    \label{fig:final_training_loss}
\end{figure}

Fig.~\ref{fig:training_loss} depicts the loss throughout the training procedure for both the linear and log-linear random walks. 
In all cases the loss quickly converges to a final value that it fluctuates around due to the random walk variability. 
The final value decreases with increasing number of qubits $n$, as shown in Fig.~\ref{fig:final_training_loss}.
As $n$ increases, the loss approaches the minimal value of $-1$, which would imply perfect correlation between the true distance to the identity and the learned guidance function. 
This could be interpreted as paradoxically implying that the problem becomes easier as $n$ grows, which certainly is not the case. 
Rather, this phenomenon can be attributed to the rapid growth of the Clifford group and the fact that the batch size $N_B$ has been kept constant as $n$ varies. 
This causes the batches to become more heterogeneous as $n$ grows, and thus the problem of learning a well-correlated guidance function on the batch becomes easier. 

Sec.~\ref{sec:results} in the main body of the text contains a detailed discussion of the results for the beam search graph traversal algorithm for the log-linear random walk, including  Table~\ref{table:comparison_beam_loglinear} which contained statistics comparing the algorithm's performance to the Qiskit method.
Analogous results for the greedy graph traversal and the linear random walk beam search are detailed in Tables~\ref{table:comparison_beam_linear}, \ref{table:comparison_greedy_loglinear}, \ref{table:comparison_greedy_linear}.

\begin{table*}[htbp]
  \centering
  \caption{\label{table:comparison_beam_linear} Beam search comparison with built-in Qiskit method (linear scaling).
  }
  \begin{tabular}{ccccc}
    \toprule
    $n$ & success (\%) & $\ell_{\text{Beam}} < \ell_{\text{Qiskit}}$ (\%) & $\ell_{\text{Beam}} \le \ell_{\text{Qiskit}}$ (\%) & $\frac{\ell_{\text{Qiskit}} - \ell_{\text{Beam}}}{\ell_{\text{Qiskit}}}  $ (\%) \\
    \midrule
             3 & 100.0 & 0 & 63.7 & N/A \\
             4 & 100.0 & 42.4 & 96.2 & 26.4 \\
             5 & 100.0 & 48.4 & 96.1 & 22.6 \\
             6 & 18.9 & 25.3 & 99.5 & 22.2 \\
             7 & 12.0 & 22.6 & 99.5 & 22.1 \\
             8 & 8.3 & 14.7 & 100.0 & 20.7 \\
             9 & 8.3 & 9.9 & 100.0 & 20.6 \\
             10 & 5.4 & 8.5 & 99.6 & 23.5 \\
             11 & 6.1 & 7.9 & 100.0 & 21.4 \\
             12 & 5.6 & 9.2 & 100.0 & 19.8 \\
    \bottomrule
  \end{tabular}
\end{table*}

\begin{table*}[htbp]
  \centering
  \caption{\label{table:comparison_greedy_loglinear} Greedy comparison with built-in Qiskit method (log-linear scaling).
  }
  \begin{tabular}{ccccc}
    \toprule
    $n$ & success (\%) & $\ell_{\text{Greedy}} < \ell_{\text{Qiskit}}$ (\%) & $\ell_{\text{Greedy}} \le \ell_{\text{Qiskit}}$ (\%) & $ \frac{\ell_{\text{Qiskit}} - \ell_{\text{Greedy}}}{\ell_{\text{Qiskit}}}  $ (\%) \\
    \midrule
   		3 & 100.0 & 0 & 38.0 & N/A \\
                 4 & 100.0 & 27.8 & 57.6 & 23.9 \\    
                 5 & 100.0 & 28.6 & 50.7 & 20.3 \\    
                 6 & 100.0 & 19.8 & 34.0 & 19.0 \\    
                 7 & 99.1 & 4.2 & 9.9 & 18.1 \\    
                 8 & 3.4 & 15.3 & 96.2 & 19.6 \\    
                 9 & 2.6 & 10.2 & 98.5 & 24.4 \\    
                 10 & 2.1 & 7.6 & 97.9 & 17.4 \\    
                 11 & 1.1 & 3.6 & 100.0 & 20.1 \\    
                 12 & 0.7 & 0.0 & 100.0 & N/A \\    
    \bottomrule
  \end{tabular}
\end{table*}

\begin{table*}[htbp]
  \centering
  \caption{\label{table:comparison_greedy_linear} Greedy comparison with built-in Qiskit method (linear scaling).
  }
  \begin{tabular}{ccccc}
    \toprule
   $n$ & success (\%) & $\ell_{\text{Greedy}} < \ell_{\text{Qiskit}}$ (\%) & $\ell_{\text{Greedy}} \le \ell_{\text{Qiskit}}$ (\%) & $ \frac{\ell_{\text{Qiskit}} - \ell_{\text{Greedy}}}{\ell_{\text{Qiskit}}}  $ (\%) \\
    \midrule
    	     3 & 100.0 & 0 & 47.5 & N/A \\
             4 & 100.0 & 18.7 & 57.3 & 25.9 \\
             5 & 43.3 & 24.3 & 80.6 & 22.6 \\
             6 & 14.1 & 16.6 & 99.5 & 21.1 \\
             7 & 10.3 & 16.1 & 99.4 & 21.9 \\
             8 & 7.2 & 8.8 & 99.9 & 20.0 \\
             9 & 6.3 & 7.6 & 99.8 & 21.9 \\
             10 & 5.4 & 7.7 & 99.7 & 22.5 \\
             11 & 4.9 & 5.2 & 100.0 & 19.2 \\
             12 & 5.0 & 5.5 & 99.4 & 20.8 \\
    \bottomrule
  \end{tabular}
\end{table*}

\end{document}